\documentclass[a4paper,11pt]{article}
\usepackage{pos}

\newcommand{\be}{\begin{equation}}
\newcommand{\ee}{\end{equation}}
\newcommand{\bea}{\begin{eqnarray}}
\newcommand{\eea}{\end{eqnarray}}
\newcommand{\MSbar}{{\overline{\rm MS}}}
\usepackage{IEEEtrantools}

\title{Gauge-invariant renormalization of fermion bilinears and energy-momentum tensor on the lattice}
\ShortTitle{Gauge-invariant renormalization of fermion bilinears and EMT on the lattice}

\author*[a,b]{G.~Spanoudes}
\author[a]{M.~Costa}
\author[a,b]{I.~Karpasitis}
\author[a,c]{T.~Pafitis}
\author[d]{G.~Panagopoulos}
\author[a]{H.~Panagopoulos}
\author[a]{A.~Skouroupathis}

\affiliation[a]{Department of Physics, University of Cyprus, Nicosia, CY-1678, Cyprus}
\affiliation[b]{Present Address: Computation-based Science and Technology Research Center, The Cyprus Institute, 20 Kavafi Str., Nicosia 2121, Cyprus}
\affiliation[c]{Present Address: Department of Physics, Utrecht University, 3508 TC Utrecht, the Netherlands}
\affiliation[d]{Department of Physics, Stanford University, California, 94305–2004, USA}

\emailAdd{g.spanoudis@cyi.ac.cy}
\emailAdd{kosta.marios@ucy.ac.cy}
\emailAdd{i.karpasitis@cyi.ac.cy}
\emailAdd{pafitis.theodosis@ucy.ac.cy}
\emailAdd{gpanago@stanford.edu}
\emailAdd{haris@ucy.ac.cy}
\emailAdd{askour02@ucy.ac.cy}

\abstract{We study a gauge-invariant renormalization scheme (GIRS) for composite operators, regularized on the lattice, by extending the coordinate space (X-space) scheme proposed some years ago. In this scheme, Green's functions of products of gauge-invariant operators located at different spacetime points are considered. Due to the gauge-invariant nature of GIRS, gauge fixing is not needed in the lattice simulations. Also, when operator mixing occurs, the gauge-variant operators (BRST variations and operators which vanish by the equations of motion) can be safely excluded from the renormalization process.

  We propose a number of variants of GIRS, including integration over time slices of the operator insertion point in a Green's function, which may lead to reduced statistical noise in lattice simulations. We employ these variants in the renormalization of fermion bilinear operators and the study of mixing between the gluon and quark energy-momentum tensor operators. We extract the one-loop conversion factors relating the nonperturbative renormalization factors in different versions of GIRS to the reference scheme of $\MSbar$. }

\FullConference{%
 The 38th International Symposium on Lattice Field Theory, LATTICE2021
  26th-30th July, 2021
  Zoom/Gather@Massachusetts Institute of Technology
}

\begin{document}
\maketitle

\section{Introduction}

Renormalization of composite operators is essential when studying matrix elements and correlation functions in Hadronic Physics. It relates bare quantities of the theory to physical ones. In order to extract nonperturbative physical results from numerical simulations on the lattice, the construction of an appropriate nonperturbative renormalization scheme is needed. For comparing lattice results to experimental data, the reference scheme of $\MSbar$ is typically employed. Since $\MSbar$ is defined in dimensional regularization using continuum perturbation theory, a direct nonperturbative renormalization of lattice correlation functions to the $\MSbar$ scheme is not feasible. However, an indirect way of taking nonperturbative results in $\MSbar$ can be achieved by using an appropriate intermediate scheme, which is applicable in both continuum and lattice: The renormalization factors which convert lattice data to the intermediate scheme can be calculated nonperturbatively on the lattice, while the conversion from the intermediate to the $\MSbar$ scheme can be implemented by calculating regularization-independent conversion factors in dimensional regularization up to some perturbative order. Some popular intermediate schemes are the RI$'$/MOM and the Schr{\"o}dinger functional schemes. In our study, we focus on the construction of an alternative intermediate scheme, referred to as gauge-invariant renormalization scheme (GIRS), which is more appropriate when operator mixing occurs.

GIRS involves correlation functions of gauge-invariant operators in coordinate space, e.g.,
\begin{equation}
  \langle \mathcal{O}_1 (x) \mathcal{O}_2 (y) \rangle, \quad (x \neq y).
  \label{GF}
  \end{equation}
  The operators $\mathcal{O}_1 (x)$, $\mathcal{O}_2 (y)$ are located at different spacetime points in order to avoid contact singularities. The main idea of this prescription is to impose regularization-independent conditions on such correlation functions in the chiral limit, similar to the RI$'$/MOM scheme, e.g.,
   \begin{equation}
     Z_{\mathcal{O}_1} \ Z_{\mathcal{O}_2} \ \langle \mathcal{O}_1 (x) \mathcal{O}_2 (y) \rangle \Big|_{x-y = \bar{z}} = \langle \mathcal{O}_1 (x) \mathcal{O}_2 (y) \rangle^{\rm tree} \Big|_{x-y = \bar{z}},
    \end{equation}
where $\bar{z}$ is the renormalization 4-vector scale [$\bar{z} \neq (0,0,0,0)$]. Older investigations of such coordinate-space renormalization prescriptions can be found, e.g., in Refs.~\cite{Gimenez:2004me,Chetyrkin:2010dx,Cichy:2012is,Cichy:2016qpu,Tomii:2018zix}. This work is a continuation of the previous studies, including a number of extensions in order to deal properly with the error in nonperturbative calculations and, most importantly, with operator mixing.

There is a number of advantages of this scheme which make easier its implementation in the lattice simulations: \\
1. Due to the gauge-independent nature of GIRS, gauge fixing is not needed; thus, the problem of Gribov copies in lattice simulations is avoided within GIRS. \\
2. Gauge-variant (GV) operators have vanishing correlation functions in GIRS; thus, when mixing occurs, all GV operators [Becchi-Rouet-Stora-Tyutin (BRST) variations and operators which vanish by the equations of motion], which can mix with gauge-invariant operators, can be safely excluded from the renormalization procedure, leading to a reduced set of mixing operators. \\
3. Contact terms are automatically excluded by the definition of the GIRS correlation functions. \\
4. Perturbative matching of GIRS and $\MSbar$ scheme is possible at high perturbative order (in most cases). Given that GIRS is defined in the massless limit, well-known techniques for calculating Feynman diagrams to very high perturbative order can be used (even in the presence of mixing).

On the other hand, there are some practical issues and challenges of GIRS: \\
1. Computations in GIRS, at a given order in perturbation theory, involve diagrams with one more loop. \\
2. When mixing occurs, the calculation of some three-point functions may be unavoidable, which are typically more noisy in simulations. \\
3. The main challenge is to find a renormalization window: $a \ll |\bar{z}| \ll \Lambda_{\rm QCD}^{-1}$ (where $a$ is the lattice spacing and $\Lambda_{\rm QCD}$ is the QCD physical scale), which must be wide enough in order to keep lattice artifacts under control and at the same time to ensure reliability of continuum perturbation theory. Improvements on the size of the window are possible via subtractions of leading-order lattice artifacts~\cite{Gimenez:2004me}, step-scaling techniques~\cite{Cichy:2016qpu}, or averaging over operator positions (see, e.g.,~\cite{Tomii:2018zix}).

A promising extension of GIRS, which is employed in the present study, is to integrate, or sum on the lattice, over time slices of the operator insertion points, while setting the time separations equal to a nonzero scale, e.g.\footnote{Without loss of generality, we set $x = (\vec{x},t)$ and $y = (\vec{0},0)$.},
 \begin{equation}
   Z_{\mathcal{O}_1} \ Z_{\mathcal{O}_2} \ \int d^3 \vec{x} \ \langle \mathcal{O}_1 (\vec{x},t) \mathcal{O}_2 (\vec{0},0) \rangle \Big|_{t = \bar{t}} = \int d^3 \vec{x} \ \langle \mathcal{O}_1 (\vec{x},t) \mathcal{O}_2 (\vec{0},0) \rangle^{\rm tree} \Big|_{t = \bar{t}},
 \label{cond}
 \end{equation}
 where $\bar{t} \neq 0$. We call this variant t-GIRS. Given the summations over time slices, the nonperturbative data in lattice simulations are expected to show small statistical errors.

 In what follows, we consider two applications of GIRS: in the multiplicative renormalization of fermion bilinear operators (Sec.~\ref{secII}) and in the study of mixing of the QCD traceless gluonic and fermionic energy-momentum tensor (EMT) operators (Sec.~\ref{secIII}). A long write-up of our work, together with an extended list of references, can be found in Ref.~\cite{Costa:2021iyv}. 

\section{Application of GIRS to the fermion bilinears}
\label{secII}

We first present our one-loop calculation for the renormalization of local fermion bilinear operators in GIRS. This computation served mostly as a prototype for the more demanding computation of the energy-momentum tensor renormalization, which follows. We consider all possible types of fermion bilinears which have definite behavior under Lorentz and parity transformations: $\mathcal{O}_X (x) = \bar{\psi} (x) X \psi (x)$, where $X = \mathbf{1}$ (scalar), $\gamma_5$ (psedoscalar), $\gamma_{\mu}$ (vector), $\gamma_5 \gamma_{\mu}$ (axial vector), and $\sigma_{\mu \nu} \equiv [\gamma_\mu, \gamma_\nu] / 2$ (tensor). One may consider both flavor singlet ($\frac{1}{N_f}\sum_f \bar \psi_f X \psi_f$) and non\-singlet operators ($\bar \psi_f X \psi_{f'}$, $f \neq f'$). For our one-loop computation, the flavor content is irrelevant and thus, we have omitted flavor indices on $\psi$, $\bar \psi$. Also, in order to avoid the mixing of flavor singlet scalar operator with the unit operator, we consider normal-ordered operators.

\begin{figure}
\hspace{0.9cm} \includegraphics[width=.19\textwidth]{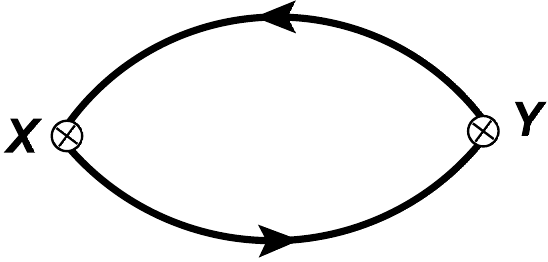} \hspace{0.6cm}
\includegraphics[width=.56\textwidth]{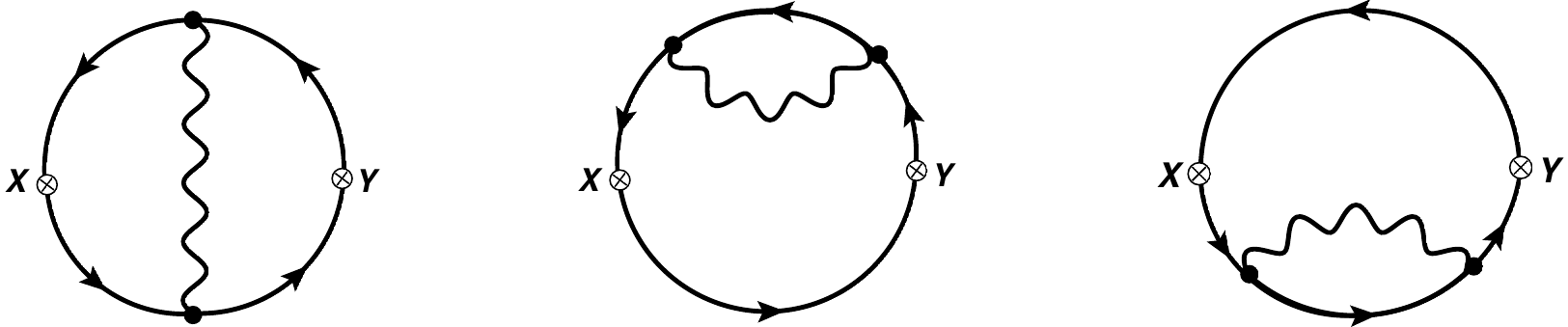}
\caption{Feynman diagrams contributing to $\langle {\cal O}_{X}(x) {\cal O}_{Y}(y)\rangle$ to order $\mathcal{O} (g^0)$ (first diagram from the left) and $\mathcal{O} (g^2)$ (three remaining diagrams). A wavy (solid) line represents gluons (quarks). A cross denotes insertion of a fermion bilinear operator. }
\label{fig:bilinears}
\end{figure}

We calculate two-point Green's functions of the form of Eq.~\ref{GF} with $\mathcal{O}_1 = \mathcal{O}_X$ and $\mathcal{O}_2 = \mathcal{O}_Y$, where $X, Y$ denote products of Dirac matrices (see definition of $X$ above). $X$ can, in principle, differ from $Y$. Note that in order to obtain a nonzero result the flavor of the fermion (antifermion) field in ${\cal O}_X$ must coincide with the flavor of antifermion (fermion) field in ${\cal O}_Y$. The Feynman diagrams contributing to these two-point functions, up to $\mathcal{O} (g^2)$, are shown in Fig.~\ref{fig:bilinears}.

The resulting expressions for all the nonvanishing $\MSbar$-renormalized correlation functions of fermion bilinears are given below:
  \begin{IEEEeqnarray}{rcl}
\langle {\cal O}^{{\overline{\rm MS}}}_{\mathbf{1}}(x) \ {{\cal O}^{{\overline{\rm MS}}}_{\mathbf{1}}} (0) \rangle &=& \frac{N_c}{\pi^{4} \ (x^2)^{3}} \Bigg[ 1 + \frac{g_{\overline{\rm MS}}^2 \ C_F}{16 \pi^2} \left(2 + 6 \ln (\bar \mu ^2 x^2) + 12 \gamma_E - 12 \ln(2) \right) \Bigg] + \ \mathcal{O} (g_{\overline{\rm MS}}^4), \\
\langle {\cal O}^{{\overline{\rm MS}}}_{\gamma_5}(x) \ {{\cal O}^{{\overline{\rm MS}}}_{\gamma_5}} (0) \rangle &=& -\langle {\cal O}^{{\overline{\rm MS}}}_{\mathbf{1}}(x) {{\cal O}^{{\overline{\rm MS}}}_{\mathbf{1}}} (0) \rangle - 16 \ c_{\rm HV} \, \frac{N_c}{\pi^{4} \ (x^2)^{3}} \frac{g_{\overline{\rm MS}}^2 \ C_F}{ 16 \pi^2} \ + \ \mathcal{O} (g_{\overline{\rm MS}}^4), \\
\langle {\cal O}^{{\overline{\rm MS}}}_{\gamma_{\mu}}(x) \ {{\cal O}^{{\overline{\rm MS}}}_{\gamma_{\nu}}} (0) \rangle &=& -\frac{N_c \ \ell_{\mu\nu}^{[1]}}{\pi^{4} \ (x^2)^{3}} \left( 1 + 3 \frac{g_{\overline{\rm MS}}^2 \ C_F}{16 \pi^2}\right) + \ \mathcal{O} (g_{\overline{\rm MS}}^4), \\
\langle {\cal O}^{{\overline{\rm MS}}}_{\gamma_5 \gamma_{\mu}}(x) \ {{\cal O}^{{\overline{\rm MS}}}_{\gamma_5 \gamma_{\nu}}} (0) \rangle &=& \langle {\cal O}^{{\overline{\rm MS}}}_{\gamma_{\mu}}(x) {{\cal O}^{{\overline{\rm MS}}}_{\gamma_{\nu}}} (0) \rangle + 8 \ c_{\rm HV} \, \frac{N_c \ \ell_{\mu\nu}^{[1]}}{\pi^{4} \ (x^2)^{3}} \frac{g_{\overline{\rm MS}}^2 \ C_F}{16 \pi^2} + \ \mathcal{O} (g_{\overline{\rm MS}}^4), \\
\langle {\cal O}^{{\overline{\rm MS}}}_{\sigma_{\mu \nu}}(x) \ {{\cal O}^{{\overline{\rm MS}}}_{\sigma_{\rho \sigma}}} (0) \rangle &=& -\frac{N_c \ \ell_{\mu\nu\rho\sigma}^{[2]}}{\pi^{4} \ (x^2)^{3}} \Bigg[ 1 + \frac{g_{\overline{\rm MS}}^2 \ C_F}{ 16 \pi^2} \left(6 - 2 \ln (\bar \mu^2 x^2)-4\gamma_E  + 4 \ln (2) \right) \Bigg] + \ \mathcal{O} (g_{\overline{\rm MS}}^4), \ \ \
 \end{IEEEeqnarray}
 where \\
 $\ell_{\mu\nu}^{[1]} \equiv \delta_{\mu \nu} - 2 \frac{x_{\mu} \ x_{\nu}}{x^2}$, \ $\ell_{\mu\nu\rho\sigma}^{[2]} \equiv (\delta_{\mu \rho} \ \delta_{\nu \sigma} - \delta_{\mu \sigma} \ \delta_{\nu \rho}) - 2 \ (\delta_{\mu \rho} \frac{x_{\nu} \ x_{\sigma}}{x^2} - \delta_{\mu \sigma} \frac{x_{\nu} \ x_{\rho}}{x^2} - \delta_{\nu \rho} \ \frac{x_{\mu} \ x_{\sigma}}{x^2} + \delta_{\nu \sigma} \frac{x_{\mu} \ x_{\rho}}{x^2})$, $C_F \equiv (N_c^2 - 1)/ (2 N_c)$, $c_{\rm HV} = 0 \ (1)$ for the naive dimensional regularization (t'Hooft-Veltman) prescription of $\gamma_5$ and the scale $\bar{\mu}$ comes from the renormalization of coupling constant in $d$ dimensions: $g_R = {(\sqrt{e^{\gamma_E}/ (4 \pi)} \ \bar{\mu})}^{(d-4)/2} Z_g^{-1} g$.  

By applying the condition of Eq.~\eqref{cond}, we extract the regularization-independent conversion factors between t-GIRS and $\MSbar$: $C_{\mathcal{O}_X}^{{\rm t-GIRS},\overline{\rm MS}} \equiv Z_{\mathcal{O}_X}^{{\rm DR}, \overline{\rm MS}} / Z_{\mathcal{O}_X}^{{\rm DR}, {\rm t-GIRS}} = Z_{\mathcal{O}_X}^{{\rm L}, \overline{\rm MS}} / Z_{\mathcal{O}_X}^{{\rm L}, {\rm t-GIRS}}$, where ${\rm DR \ (L)}$ denotes dimensional (lattice) regularization. The one-loop results for the nonvanishing cases are:
  \begin{IEEEeqnarray}{rcl}
    C_{\mathcal{O}_\mathbf{1}}^{{\rm t-GIRS},{\overline{\rm MS}}} &=& 1 \ + \ \frac{g_{\overline{\rm MS}}^2 \ C_F}{16 \pi^2} \left(-\frac{1}{2} + 6 \ln (\bar{\mu} \ \bar{t}) +6 \gamma_E\right) + \mathcal{O} (g_{\overline{\rm MS}}^4), \\
    C_{\mathcal{O}_{\gamma_5}}^{{\rm t-GIRS},{\overline{\rm MS}}} &=& 1 \ + \ \frac{g_{\overline{\rm MS}}^2 \ C_F}{16 \pi^2}\left(-\frac{1}{2} + 6 \ln (\bar{\mu} \ \bar{t}) + 6 \gamma_E + 8 \, c_{\rm HV} \right) + \mathcal{O} (g_{\overline{\rm MS}}^4), \\
    C_{\mathcal{O}_{\gamma_i}}^{{\rm t-GIRS},{\overline{\rm MS}}} &=& 1 \ + \ \frac{g_{\overline{\rm MS}}^2 \ C_F}{16 \pi^2} \left(\frac{3}{2}\right) + \mathcal{O} (g_{\overline{\rm MS}}^4), \\
    C_{\mathcal{O}_{\gamma_5 \gamma_i}}^{{\rm t-GIRS},{\overline{\rm MS}}} &=& 1 \ + \ \frac{g_{\overline{\rm MS}}^2 \ C_F}{16 \pi^2} \left(\frac{3}{2} + 4\,c_{\rm HV}\right) + \mathcal{O} (g_{\overline{\rm MS}}^4), \\
  C_{\mathcal{O}_{\sigma_{\mu \nu}}}^{{\rm t-GIRS},{\overline{\rm MS}}} &=& 1 \ + \ \frac{g_{\overline{\rm MS}}^2 C_F}{16 \pi^2}\left(\frac{25}{6} -2 \ln(\bar \mu \ \bar{t} ) -2 \gamma_E \right) + \mathcal{O} (g_{\overline{\rm MS}}^4).  
 \end{IEEEeqnarray}                                   
  
\section{Application of GIRS to the QCD traceless energy-momentum tensor}
\label{secIII}

In this section, we present our one-loop calculation for the renormalization of QCD traceless energy-momentum tensor (EMT) operators in GIRS. EMT is the conserved current associated with the spacetime translational symmetry. It is decomposed into traceless and trace parts, gluonic and fermionic components. The nucleon matrix elements of its traceless (gluonic and fermionic) components enter the nucleon spin decomposition, and they are directly related to the gluon and quark average momentum fraction of a nucleon state. While EMT is a finite quantity, the individual components are not. In our study, we focus on the renormalization of each traceless part.

The traceless (gluonic and fermionic) EMT operators are defined as:
\begin{equation}
    \overline{T}^{G}_{\mu \nu} (x) = -2 {\rm Tr} [F_{\rho \{\mu} (x) \ F_{\nu \} \rho} (x)], \hspace{1.3cm} \overline{T}^{F}_{\mu \nu} (x) = \sum_{f=1}^{N_f} \bar{\psi}_f (x) \gamma_{\{\mu} \overleftrightarrow{D}_{\nu \}} \psi_f (x),
  \end{equation}
  where $\overleftrightarrow{D}_\mu$ is the symmetrized covariant derivative and $\{ \ldots \}$ denotes the symmetrization over Lorentz indices $\mu, \nu$ and subtraction of the trace\footnote{For the explicit definitions, see Ref.\cite{Costa:2021iyv}.}. A difficulty in studying the renormalization of these operators is that mixing is present; the two operators along with three gauge-variant operators (two BRST variations and one operator vanishing by the equations of motion)~\cite{Caracciolo:1991cp}:
  \begin{eqnarray}
    \mathcal{O}_{{\rm BRST}_1} (x) &=& \frac{4}{\alpha} {\rm Tr} [ \partial_{\{ \mu} A_{\nu \}} (x) \partial_\rho A_\rho (x) - \bar{c} (x) \partial_{\{ \mu} D_{\nu \}} c (x)], \\
    \mathcal{O}_{{\rm BRST}_2} (x) &=& -\frac{4}{\alpha} {\rm Tr} [ A_{\{ \mu} (x) \partial_{\nu \}} \partial_\rho A_\rho (x) - \partial_{\{ \mu} \bar{c} (x) D_{\nu \}} c (x)], \\
    \mathcal{O}_{\rm EOM} (x) &=& 4 {\rm Tr} [ A_{\{ \mu} (x) \ \delta S / \delta A_{\nu \}} (x)],
  \end{eqnarray}
mix among themselves, as they have the same transformations under Euclidean rotational (or hypercubic, on the lattice) symmetry. The three GV operators depend on ghost fields and gauge-fixing terms, which are well-defined in perturbation theory, while their nonperturbative extensions are not obvious; thus, a nonperturbative study of such terms by compact lattice simulations is nontrivial. For a recent two-loop calculation of all mixing coefficients in dimensional regularization, see our Ref.~\cite{Panagopoulos:2020qcn}. However, employing GIRS the GV operators are automatically excluded from the renormalization process. Then, for the renormalization of the two remaining operators (the two EMT operators), we need to construct a $2 \times 2$ mixing matrix:
\begin{equation}
\begin{pmatrix}
\overline{T}^{G \ R}_{\mu \nu} \\
\overline{T}^{F \ R}_{\mu \nu}
 \end{pmatrix} =
 \begin{pmatrix}
Z_{GG} & Z_{GF} \\
Z_{FG} & Z_{FF}
 \end{pmatrix}
 \begin{pmatrix}
\overline{T}^G_{\mu \nu} \\
\overline{T}^F_{\mu \nu}.
 \end{pmatrix}
\end{equation}

The calculation of all mixing matrix elements requires a total of four conditions involving correlation functions of $\overline{T}^G_{\mu \nu}$ and $\overline{T}^F_{\mu \nu}$. Three conditions can be obtained by considering two-point functions between the two EMT operators:
 \begin{eqnarray}
      \langle \overline{T}^{G \ {\rm GIRS}}_{\mu \nu} (x) \overline{T}^{G \ {\rm GIRS}}_{\rho \sigma} (y) \rangle \Big|_{x-y=\bar{z}} &=& {\langle \overline{T}^G_{\mu \nu} (x) \overline{T}^G_{\rho \sigma} (y) \rangle}^{\rm tree} \Big|_{x-y=\bar{z}} \ , \\
      \langle \overline{T}^{F \ {\rm GIRS}}_{\mu \nu} (x) \overline{T}^{F \ {\rm GIRS}}_{\rho \sigma} (y) \rangle \Big|_{x-y=\bar{z}} &=& {\langle \overline{T}^F_{\mu \nu} (x) \overline{T}^F_{\rho \sigma} (y) \rangle}^{\rm tree} \Big|_{x-y=\bar{z}} \ , \\
      \langle \overline{T}^{G \ {\rm GIRS}}_{\mu \nu} (x) \overline{T}^{F \ {\rm GIRS}}_{\rho \sigma} (y) \rangle \Big|_{x-y=\bar{z}} &=& {\langle \overline{T}^G_{\mu \nu} (x) \overline{T}^F_{\rho \sigma} (y) \rangle}^{\rm tree} \Big|_{x-y=\bar{z}} = 0.
    \end{eqnarray}
    A fourth condition can be obtained by considering three-point functions among an EMT operator and two lower dimensional operators, e.g., two fermion bilinears~\footnote{Two-point functions between an EMT operator and one fermion bilinear operator vanish due to trace algebra or charge conjugation symmetry.}:
    \begin{equation}
      \langle \mathcal{O}_X^{\rm GIRS} (x) \overline{T}^{G \ {\rm GIRS}}_{\mu \nu} (w) \mathcal{O}_Y^{\rm GIRS} (y) \rangle \Big|_{\begin{smallmatrix}
          x-w = \bar{z}, \\
        y-w = \bar{z}', \\
          \bar{z} \neq \bar{z}'
        \end{smallmatrix}} = {\langle \mathcal{O}_X (x) \overline{T}^G_{\mu \nu} (w) \mathcal{O}_Y (y) \rangle}^{\rm tree} \Big|_{\begin{smallmatrix}
          x-w = \bar{z}, \\
        y-w = \bar{z}', \\
        \bar{z} \neq \bar{z}'
        \end{smallmatrix}} = 0.
      \end{equation} 
For simplifying the perturbative calculation, we choose $\bar{z}'=-\bar{z}$. Also, we employ the offdiagonal $(\mu \neq \nu)$ elements of EMT and flavor nonsinglet fermion bilinears. Different extensions of our calculation, including the diagonal components, are under investigation.   

Similarly, we define renormalization conditions in t-GIRS. As an example, we employ two vector bilinear operators; we choose the free Lorentz indices in such a way as to obtain a solvable system of equations:
\begin{eqnarray}
      \int d^3 \vec{x} \ \langle \overline{T}^{G \ {\rm t-GIRS}}_{ij} (\vec{x},t) \overline{T}^{G \ {\rm t-GIRS}}_{ij} (\vec{0},0) \rangle |_{t=\bar{t}} &=& {\rm tree}, \quad (i \neq j) \label{condEMT1} \\
      \int d^3 \vec{x} \ \langle \overline{T}^{F \ {\rm t-GIRS}}_{ij} (\vec{x},t) \overline{T}^{F \ {\rm t-GIRS}}_{ij} (\vec{0},0) \rangle |_{t=\bar{t}} &=& {\rm tree}, \quad (i \neq j) \\
      \int d^3 \vec{x} \ \langle \overline{T}^{G \ {\rm t-GIRS}}_{ij} (\vec{x},t) \overline{T}^{F \ {\rm t-GIRS}}_{ij} (\vec{0},0) \rangle |_{t=\bar{t}} &=& {\rm tree}, \quad (i \neq j) \\
      \int d^3 \vec{x} \ \langle \mathcal{O}_{\gamma_i}^{\rm t-GIRS} (\vec{x},t) \overline{T}^{G \ {\rm t-GIRS}}_{ij} (\vec{0},0) \mathcal{O}_{\gamma_j}^{\rm t-GIRS} (-\vec{x},-t) \rangle |_{t=\bar{t}} &=& {\rm tree}, \quad (i \neq j) \label{condEMT4}
    \end{eqnarray}
    where the abbreviation ``tree'' in the r.h.s. of each condition corresponds to the tree-level value of the l.h.s.

\begin{figure}
  \hspace{0.83cm} \includegraphics[width=.145\textwidth]{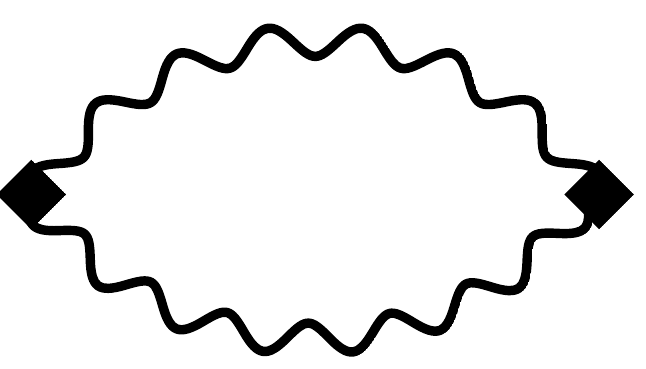} \hspace{0.5cm} \includegraphics[width=.675\textwidth]{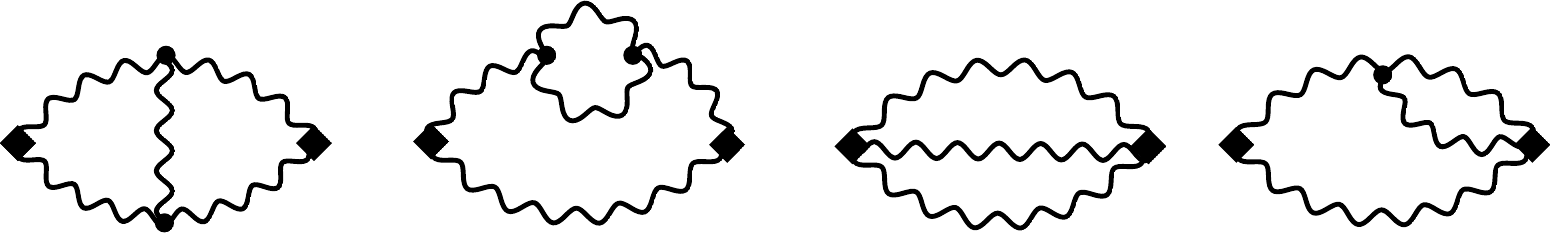} \\
\hspace*{2.13cm}\includegraphics[width=.675\textwidth]{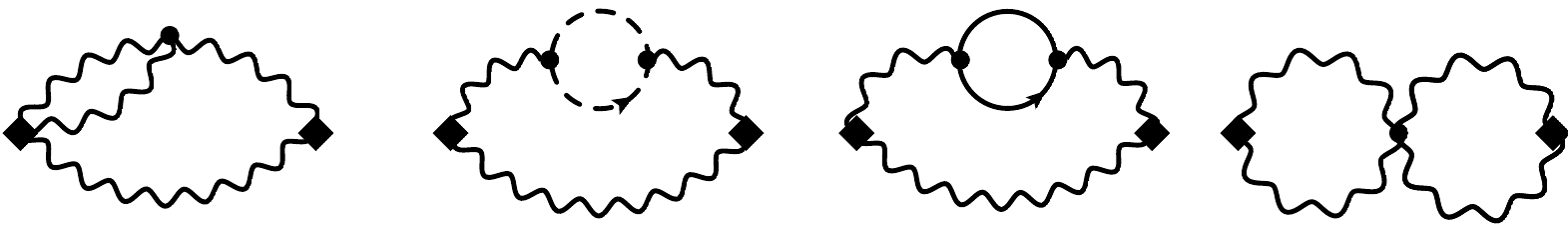}
\caption{Feynman diagrams contributing to $\langle \overline{T}^G_{\mu \nu}(x) \overline{T}^G_{\rho \sigma} (y) \rangle$, to order $\mathcal{O} (g^0)$ (first diagram (upper left)) and $\mathcal{O} (g^2)$ (the remaining diagrams). Wavy (solid, dashed) lines represent gluons (quarks, ghosts). A diamond denotes insertion of the gluon EMT operator.}
\label{fig:2ptEMTgg}
\end{figure}

\begin{figure}
  \hspace{0.5cm} \includegraphics[width=.145\textwidth]{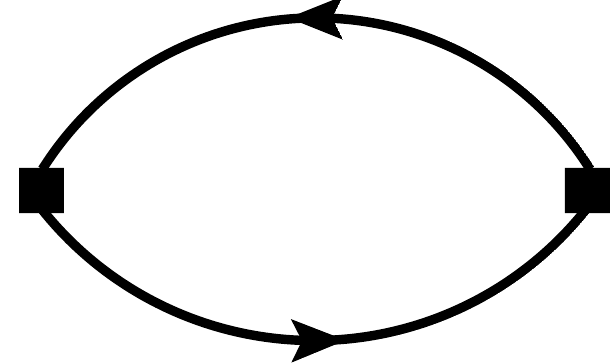} \hspace{0.5cm} \includegraphics[width=.7\textwidth]{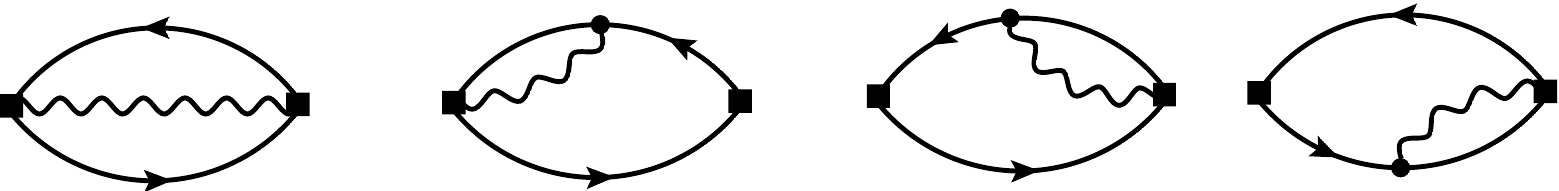} \\
\hspace*{1.8cm}\includegraphics[width=.7\textwidth]{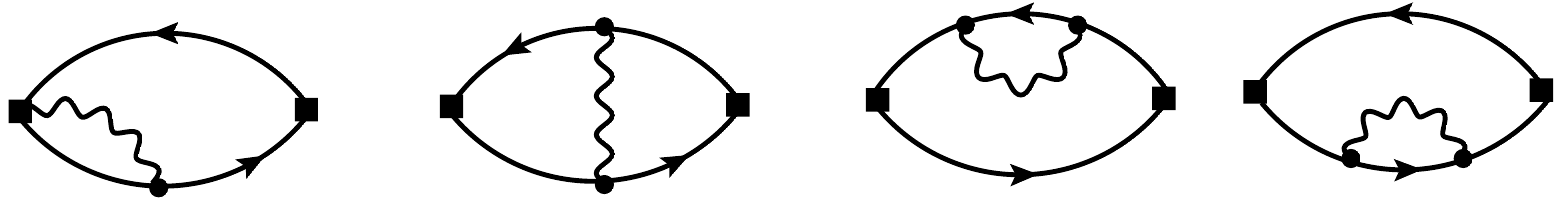}
\caption{Feynman diagrams contributing to $\langle \overline{T}^F_{\mu \nu}(x) \overline{T}^F_{\rho \sigma} (y) \rangle$, to order $\mathcal{O} (g^0)$ (first diagram (upper left)) and $\mathcal{O} (g^2)$ (the remaining diagrams). Wavy (solid) lines represent gluons (quarks). A square denotes insertion of the fermion EMT operator.}
\label{fig:2ptEMTff}
\end{figure}

\begin{figure}
\centering \includegraphics[width=.37\textwidth]{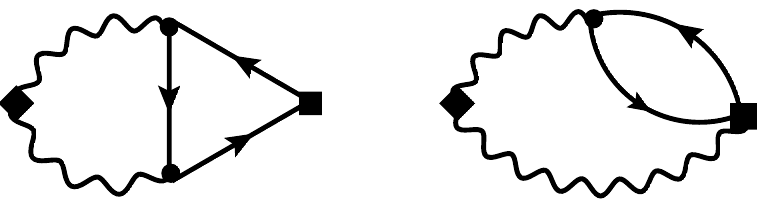} 
\caption{Feynman diagrams contributing to $\langle \overline{T}^G_{\mu \nu}(x) \overline{T}^F_{\rho \sigma} (y) \rangle$, to order $\mathcal{O} (g^2)$. There is no $\mathcal{O} (g^0)$ contribution. Wavy (solid) lines represent gluons (quarks). A diamond (square) denotes insertion of the gluon (fermion) EMT operator.}
\label{fig:2ptEMTgf}
\end{figure}

\begin{figure}
\hspace{1.2cm} \includegraphics[width=.12\textwidth]{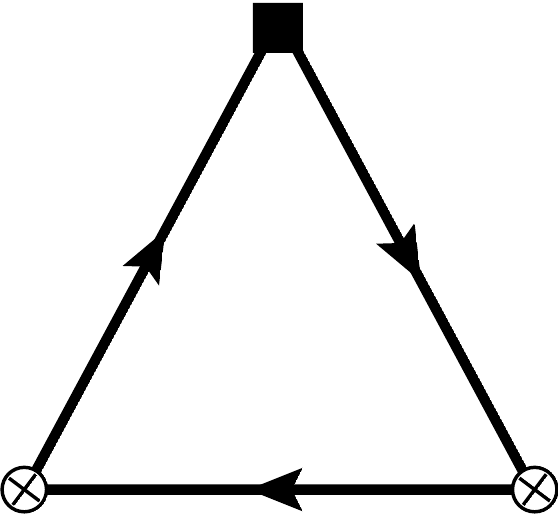} \hspace{0.4cm}
\includegraphics[width=.47\textwidth]{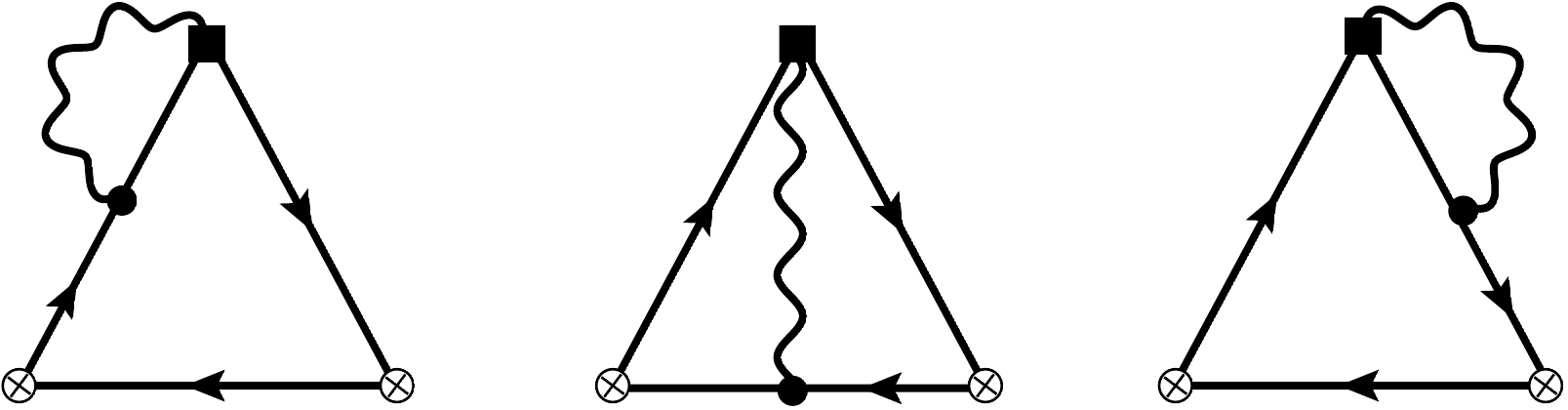} \hspace{0.4cm}
\includegraphics[width=.13\textwidth]{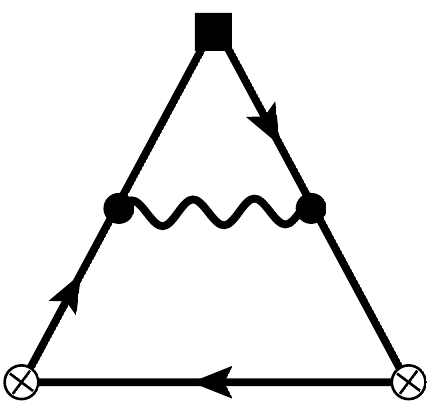} \vspace{0.3cm}\\
\hspace*{1.1cm}\includegraphics[width=.3\textwidth]{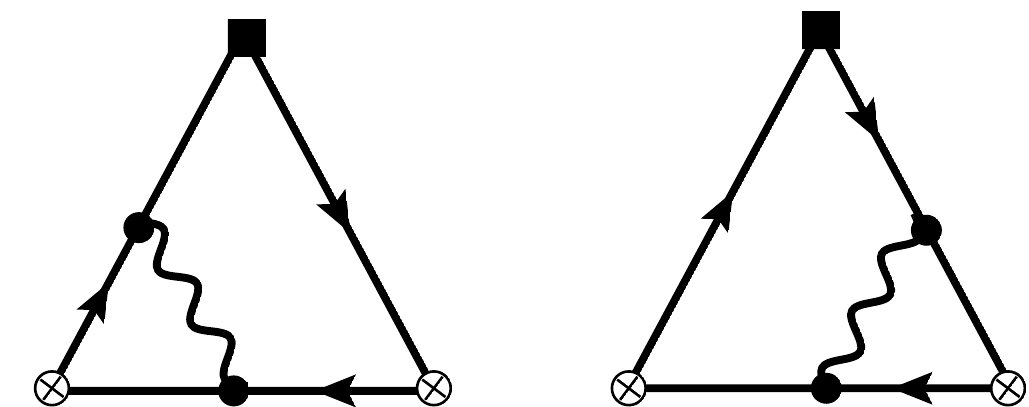} \vspace*{-1.9cm}\\
 \hspace*{6.3cm} \includegraphics[width=.48\textwidth]{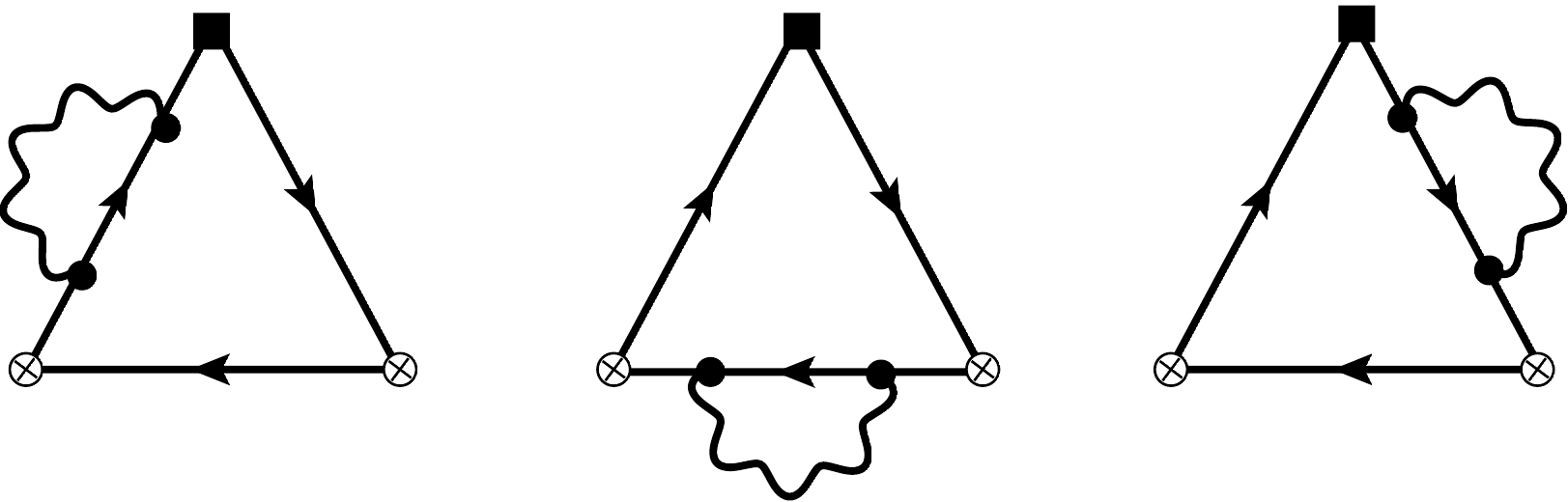}
\caption{Feynman diagrams contributing to $\langle \mathcal{O}_X (x) \overline{T}^F_{\mu \nu} (w) \mathcal{O}_Y (y) \rangle$, to order $\mathcal{O} (g^0)$ (first diagram (upper left)) and $\mathcal{O} (g^2)$ (the remaining diagrams). Wavy (solid) lines represent gluons (quarks). A square (cross) denotes insertion of the fermion EMT (bilinear) operator. Diagrams having the arrows of the fermion lines in counterclockwise direction must also be considered.}
\label{fig:3ptEMTf}
\end{figure}

\begin{figure}
\centering \includegraphics[width=.52\textwidth]{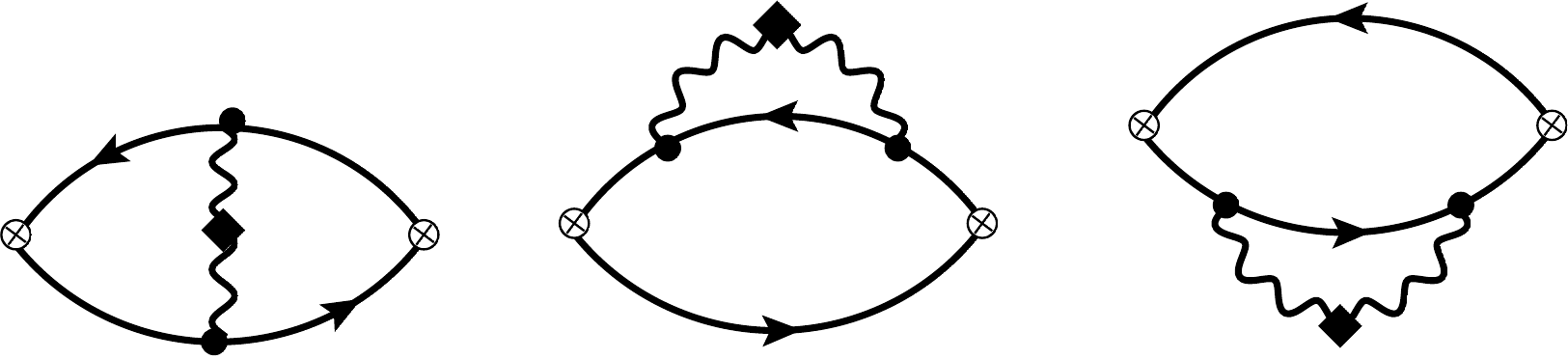} 
\caption{Feynman diagrams contributing to $\langle \mathcal{O}_X (x) \overline{T}^G_{\mu \nu} (w) \mathcal{O}_Y (y) \rangle$, to order $\mathcal{O} (g^2)$. There is no $\mathcal{O} (g^0)$ contribution. Wavy (solid) lines represent gluons (quarks). A diamond (cross) denotes insertion of the gluon EMT (fermion bilinear) operator.}
\label{fig:3ptEMTg}
\end{figure}

    The Feynman diagrams contributing to the two-point and three-point functions of the EMT operators, up to $\mathcal{O} (g^2)$, are shown in Figs. (\ref{fig:2ptEMTgg},\ref{fig:2ptEMTff},\ref{fig:2ptEMTgf},\ref{fig:3ptEMTf},\ref{fig:3ptEMTg}). The resulting expressions for the nonvanishing $\MSbar$-renormalized two-point and three-point correlation functions of EMT operators are given below, up to $\mathcal{O} (g_{\overline{\rm MS}}^2)$. Results for the three-point functions with other fermion bilinears can be found in our manuscript~\cite{Costa:2021iyv}.
\begin{IEEEeqnarray}{rcl}
\langle \overline{T}^{G \ \overline{\rm MS}}_{\mu \nu} (x) \overline{T}^{G \ \overline{\rm MS}}_{\rho \sigma} (0) \rangle &=& \frac{4 N_c C_F \ s_{\mu \nu \rho \sigma}^{[1]}}{\pi^{4} {(x^2)}^{4}} \Bigg[ 1 - \frac{g_{\overline{\rm MS}}^2}{ 16 \pi^2} \frac{4}{3} \Bigg(N_f \Big(\frac{1}{6} + \ln (\bar{\mu}^2 x^2) + 2 \gamma_E - 2 \ln (2) \Big) \nonumber \\
&& \hspace{4.5cm} + \frac{5 N_c}{3} \Bigg)\Bigg], \\
\langle \overline{T}^{F \ \overline{\rm MS}}_{\mu \nu} (x) \overline{T}^{F \ \overline{\rm MS}}_{\rho \sigma} (0) \rangle &=& \frac{N_c N_f \ s_{\mu \nu \rho \sigma}^{[1]}}{\pi^{4} {(x^2)}^{4}} \Bigg[ 1 - \frac{g_{\overline{\rm MS}}^2}{ 16 \pi^2} \frac{16 C_F}{3} \Big(-\frac{59}{48} + \ln (\bar{\mu}^2 x^2) + 2 \gamma_E - 2 \ln (2) \Big)\Bigg] \ \ \\
\langle \overline{T}^{G \ \overline{\rm MS}}_{\mu \nu} (x) \overline{T}^{F \ \overline{\rm MS}}_{\rho \sigma} (0) \rangle &=& \frac{N_c N_f \ s_{\mu \nu \rho \sigma}^{[1]}}{\pi^{4} {(x^2)}^{4}} \Bigg[ \frac{g_{\overline{\rm MS}}^2}{ 16 \pi^2} \frac{16 C_F}{3} \Big(-\frac{1}{6} + \ln (\bar{\mu}^2 x^2) + 2 \gamma_E - 2 \ln (2) \Big)\Bigg],
\end{IEEEeqnarray}
\begin{IEEEeqnarray}{rcl}
  \langle \mathcal{O}_{\gamma_\rho}^{\overline{\rm MS}} (x) \overline{T}^{G \ \overline{\rm MS}}_{\mu \nu} (0) \mathcal{O}_{\gamma_\sigma}^{\overline{\rm MS}} (-x) \rangle &=& \frac{N_c N_f}{4 \pi^{6} {(x^2)}^{5}} \frac{g_{\overline{\rm MS}}^2}{ 16 \pi^2} \frac{8 C_F}{3} \Bigg[ s_{\mu \nu \rho \sigma}^{[2]} \Big(-1.701491 + \ln (\bar{\mu}^2 x^2) \Big) + s_{\mu \nu \rho \sigma}^{[3]} \Bigg] \ \ \ \ \ \ \\
   \langle \mathcal{O}_{\gamma_\rho}^{\overline{\rm MS}} (x) \overline{T}^{F \ \overline{\rm MS}}_{\mu \nu} (0) \mathcal{O}_{\gamma_\sigma}^{\overline{\rm MS}} (-x) \rangle &=& \frac{N_c N_f}{4 \pi^{6} {(x^2)}^{5}} \Bigg[ s_{\mu \nu \rho \sigma}^{[2]} - \frac{g_{\overline{\rm MS}}^2}{ 16 \pi^2} \frac{8 C_F}{3} \Bigg( s_{\mu \nu \rho \sigma}^{[2]} \Big(-3.201491 + \ln (\bar{\mu}^2 x^2) \Big) \nonumber \\
&& \hspace{4.9cm} + s_{\mu \nu \rho \sigma}^{[4]} \Bigg) \Bigg],
 \end{IEEEeqnarray} 
 where \\
 $s_{\mu \nu \rho \sigma}^{[1]} = (\delta_{\mu \rho} \delta_{\nu \sigma} + \delta_{\mu \sigma} \delta_{\nu \rho}) + 8 \ \frac{x_\mu x_\nu x_\rho x_\sigma}{{(x^2)}^2} -2 \ (\delta_{\mu \rho} \frac{x_{\nu} \ x_{\sigma}}{x^2} + \delta_{\mu \sigma} \frac{x_{\nu} \ x_{\rho}}{x^2} + \delta_{\nu \rho} \ \frac{x_{\mu} \ x_{\sigma}}{x^2} + \delta_{\nu \sigma} \frac{x_{\mu} \ x_{\rho}}{x^2})$, \\
 $s_{\mu \nu \rho \sigma}^{[2]} = 2 \frac{x_\mu x_\nu}{x^2} \delta_{\rho \sigma} - 8 \frac{x_\mu x_\nu x_\rho x_\sigma}{{(x^2)}^2} + (\delta_{\mu \rho} \frac{x_{\nu} \ x_{\sigma}}{x^2} + \delta_{\mu \sigma} \frac{x_{\nu} \ x_{\rho}}{x^2} + \delta_{\nu \rho} \ \frac{x_{\mu} \ x_{\sigma}}{x^2} + \delta_{\nu \sigma} \frac{x_{\mu} \ x_{\rho}}{x^2})$, \\
 $s_{\mu \nu \rho \sigma}^{[3]} = \frac{1}{2} \frac{x_\mu x_\nu}{x^2} \delta_{\rho \sigma} + \frac{3}{4} (\delta_{\mu \rho} \delta_{\nu \sigma} + \delta_{\mu \sigma} \delta_{\nu \rho}) - (\delta_{\mu \rho} \frac{x_{\nu} \ x_{\sigma}}{x^2} + \delta_{\mu \sigma} \frac{x_{\nu} \ x_{\rho}}{x^2} + \delta_{\nu \rho} \ \frac{x_{\mu} \ x_{\sigma}}{x^2} + \delta_{\nu \sigma} \frac{x_{\mu} \ x_{\rho}}{x^2})$, \\
 $s_{\mu \nu \rho \sigma}^{[4]} = \frac{11}{4} \frac{x_\mu x_\nu}{x^2} \delta_{\rho \sigma} + \frac{9}{8} (\delta_{\mu \rho} \delta_{\nu \sigma} + \delta_{\mu \sigma} \delta_{\nu \rho}) - (\delta_{\mu \rho} \frac{x_{\nu} \ x_{\sigma}}{x^2} + \delta_{\mu \sigma} \frac{x_{\nu} \ x_{\rho}}{x^2} + \delta_{\nu \rho} \ \frac{x_{\mu} \ x_{\sigma}}{x^2} + \delta_{\nu \sigma} \frac{x_{\mu} \ x_{\rho}}{x^2})$. \\  

By applying the conditions of Eqs. (\ref{condEMT1} -- \ref{condEMT4}), we extract the regularization-independent conversion factors (which are now a $2 \times 2$ matrix) between t-GIRS and $\MSbar$:
 \begin{eqnarray}
\begin{pmatrix}
C_{GG}^{{\rm t-GIRS},\overline{\rm MS}} & C_{GF}^{{\rm t-GIRS},\overline{\rm MS}} \\
\\
C_{FG}^{{\rm t-GIRS},\overline{\rm MS}} & C_{FF}^{{\rm t-GIRS},\overline{\rm MS}}
\end{pmatrix} &=& \begin{pmatrix}
Z_{GG}^{{\rm DR},\overline{\rm MS}} & Z_{GF}^{{\rm DR},\overline{\rm MS}} \\
\\
Z_{FG}^{{\rm DR},\overline{\rm MS}} & Z_{FF}^{{\rm DR},\overline{\rm MS}}
\end{pmatrix} \cdot \begin{pmatrix}
Z_{GG}^{{\rm DR},{\rm t-GIRS}} & Z_{GF}^{{\rm DR},{\rm t-GIRS}} \\
\\
Z_{FG}^{{\rm DR},{\rm t-GIRS}} & Z_{FF}^{{\rm DR},{\rm t-GIRS}}
\end{pmatrix}^{-1}  \nonumber \\
  &=& \begin{pmatrix}
Z_{GG}^{{\rm L},\overline{\rm MS}} & Z_{GF}^{{\rm L},\overline{\rm MS}} \\
\\
Z_{FG}^{{\rm L},\overline{\rm MS}} & Z_{FF}^{{\rm L},\overline{\rm MS}}
\end{pmatrix} \cdot \begin{pmatrix}
Z_{GG}^{{\rm L},{\rm t-GIRS}} & Z_{GF}^{{\rm L},{\rm t-GIRS}} \\
\\
Z_{FG}^{{\rm L},{\rm t-GIRS}} & Z_{FF}^{{\rm L},{\rm t-GIRS}}
\end{pmatrix}^{-1}.
\end{eqnarray}
The one-loop results are:
    \begin{eqnarray}
      C_{GG}^{{\rm t-GIRS}, {\overline{\rm MS}}} &=& 1 - \frac{g_{\overline{\rm MS}}^2}{16 \pi^2} \Big[ \frac{10}{9} N_c + 0.236288 N_f + \frac{4}{3} N_f \ln (\bar{\mu} \ \bar{t}) \Big] + \mathcal{O} (g_{\overline{\rm MS}}^4), \\
C_{GF}^{{\rm t-GIRS}, {\overline{\rm MS}}} &=& \ \ - \frac{g_{\overline{\rm MS}}^2}{16 \pi^2} C_F \Big[-7.848365 - \frac{16}{3} \ln (\bar{\mu} \ \bar{t}) \Big] + \mathcal{O} (g_{\overline{\rm MS}}^4), \\
C_{FG}^{{\rm t-GIRS}, {\overline{\rm MS}}} &=& \ \ - \frac{g_{\overline{\rm MS}}^2}{16 \pi^2} N_f \Big[1.933961 - \frac{4}{3} \ln (\bar{\mu} \ \bar{t}) \Big] + \mathcal{O} (g_{\overline{\rm MS}}^4), \\
      C_{FF}^{{\rm t-GIRS}, {\overline{\rm MS}}} &=& 1 - \frac{g_{\overline{\rm MS}}^2}{16 \pi^2} C_F \Big[-2.777072 + \frac{16}{3} \ln (\bar{\mu} \ \bar{t}) \Big] + \mathcal{O} (g_{\overline{\rm MS}}^4).
    \end{eqnarray}
    
\section{Conclusions and future prospects}

In this work, we have studied the application of GIRS and its variant, called t-GIRS, to the renormalization of fermion bilinears and of traceless EMT operators. We have calculated the one-loop conversion factors relating t-GIRS to the $\MSbar$ scheme. The corresponding application of t-GIRS in lattice simulations is currently under investigation, including different extensions in the renormalization conditions. We have also considered some further applications of GIRS regarding the renormalization of supersymmetric operators on the lattice, such as the gluino-glue operator~\cite{Costa:2021pfu} and the supercurrent (ongoing~\cite{Georg}). Finally, a natural continuation of our work is the renormalization of the trace parts of EMT using t-GIRS, which has some further complications on the lattice.

\acknowledgments \vspace*{-0.1cm}M.C. and H.P. acknowledge financial support from the project “Quantum Fields on the Lattice”, funded by the Cyprus Research and Innovation Foundation (RIF) under contract number EXCELLENCE/0918/0066.

\bibliographystyle{JHEP}
\bibliography{LATTICE2021_GIRS_GS.bib}

\end{document}